\documentclass[11pt]{article}
\usepackage{amssymb,latexsym,amsmath}
\usepackage{amsfonts, graphics}
\newcommand{\R}{\mathbb R}
\def\open#1{\setbox0=\hbox{$#1$}
\baselineskip = 0pt
\vbox{\hbox{\hspace*{0.4 \wd0}\tiny $\circ$}\hbox{$#1$}}
\baselineskip = 11pt\!}
\def\fn{\open{f}}

\begin{document}
\title{A numerical investigation of
 the stability of steady states and critical phenomena for
the spherically symmetric Einstein-Vlasov system}
\author{H{\aa}kan Andr\'{e}asson\\
        Department of Mathematics, Chalmers,\\
        S-41296 G\"oteborg, Sweden\\
        email\textup{: \texttt{hand@math.chalmers.se}}\\
        \phantom{tom rad}\\
        Gerhard Rein\\
        Department of Mathematics, University of
        Bayreuth,\\
        D-95440 Bayreuth, Germany\\
        email\textup{: \texttt{gerhard.rein@uni-bayreuth.de}}}

\maketitle

\begin{abstract}
The stability features of steady states of the spherically
symmetric Einstein-Vlasov system are investigated numerically. We
find support for the conjecture by Zeldovich and Novikov that the
binding energy maximum along a steady state sequence signals the
onset of instability, a conjecture which we extend to and confirm 
for non-isotropic states. The sign of
the binding energy of a solution turns out to be relevant for
its time evolution in general. We relate the stability properties
to the question of universality in critical collapse and find
that for Vlasov matter universality does not seem to hold.
\end{abstract}

\section{Introduction}
For any given dynamical system the existence and stability of steady states  
is essential both from a mathematics and from a physics point of
view. In this paper we investigate these questions by numerical
means for the spherically symmetric, asymptotically flat
Einstein-Vlasov system. This system describes a self gravitating
collisionless gas in the framework of general relativity. Here the
matter is thought of as a large ensemble of particles, which is
described by a density function on phase space, and the individual
particles move along geodesics. The precise formulation of this
system will be given in the next section; for further information
on the Einstein-Vlasov system we refer to \cite{An1}. 
In astrophysics the system is used
to model compact star clusters and galaxies. In this context the
stability question was first studied by Zeldovich e.~a.\ in the
sixties \cite{ZP,ZN}. These authors characterize a steady state by
its central redshift and binding energy and conjecture that
the binding energy maximum along a steady state
sequence signals the onset of instability. 
For isotropic steady states Ipser \cite{I1} and
Shapiro and Teukolsky \cite{Sh2} found numerical support for this
conjecture. In our investigation we
find that this conjecture holds for non-isotropic steady
states as well, where the density on phase space depends on the particle
energy and angular momentum. Moreover, we use three different kinds of 
perturbations (for details see Section 3) while in the previous 
studies mentioned above only one type of perturbation was implemented.  
We also find that the sign of the binding 
energy is crucial for the evolution of the perturbed solutions. 
A positive value 
implies that the solution is bound in the sense 
that not all matter can disperse to infinity, and in this case 
the perturbation of a stable state seems to lead to periodic oscillations.
In the case of a negative value of the binding energy we observe that 
the solution disperses to infinity. 

Our initial motivation for studying
the stability of steady states was its role in critical collapse.
This topic started with the work of Choptuik \cite{Cho93} where he
studied the Einstein-scalar-field system. He took a fixed initial
profile for the scalar field and scaled it by an arbitrary
constant factor. This gives rise to a family of initial data
depending on a real parameter $A$. It turned out that there exists
a critical parameter $A_\ast$ such that for $A<A_\ast$ the
corresponding solutions disperses as predicted by the theoretical
results of Christodoulou \cite{chr86} for small data, while for
$A>A_\ast$ the corresponding solutions collapse and produce a
black hole in accordance with \cite{chr87}. The surprising result
was that the limit of the mass $M(A)$ of the black hole tends to
zero for $A \to A_\ast$ so that in such a one-parameter family
there are black holes with arbitrarily small mass. Choptuik found
that this fact is related to the existence of self-similar
solutions of the Einstein-scalar-field system, in particular,
the critical solution is self-similar and universal, i.e.,
independent of the initial profile which determines the
one-parameter family. Later on Choptuik, Chmaj, and Bizo\'n 
performed a similar investigation
for the Einstein-Yang-Mills system \cite{cho96}. 
Here both cases $\lim_{A\to
A_\ast} M(A)=0$ and $\lim_{A\to A_\ast,A >A_\ast} M(A)>0$ were
found and called type II and type I respectively. In the latter case
there is a mass gap in the $M(A)$-curve. The possibility of type I
behavior in this system was related to the existence of the
Bartnik-McKinnon solutions \cite{bar88,smo91}, which are static.
Again, the conjecture is that the critical solution is universal. 

As opposed to the field theoretic matter models mentioned above
the Vlasov model is phenomenological, and in contrast to fluid models 
several global results have been obtained for Vlasov matter. 
For the spherically symmetric, asymptotically flat case it was 
shown in \cite{RR1}
that sufficiently small initial data launch global, geodesically
complete solutions which disperse for large times. It is also
known that there do exist initial data, necessarily large, which
develop singularities \cite{Rl1}. The proof relies on the Penrose
singularity theorem. There are no general results on the behavior
of large data solutions yet, except for the following: If data on
a hypersurface of constant Schwarzschild time give rise to a
solution which develops a singularity after a finite amount of
Schwarzschild time, then the first singularity occurs at the
center of symmetry \cite{RRS1}. An analogous result where
Schwarzschild time is replaced by maximal slicing has also been
proved \cite{Rl1}. In \cite{An2} further results on the global
behavior of solutions for large data can be found. The transition
between dispersion and gravitational collapse was numerically
investigated by Rein, Rendall, and Schaeffer \cite{RRS2}, and it
was found that there is a mass gap in the $M(A)$ curve. This
result was later confirmed by Olabarrieta and Choptuik \cite{OC}.
In addition, the latter authors reported evidence that the mass gap is
due to the presence of static solutions, and they found support that 
the critical solution is universal. 

In the present investigation we address the role of steady states
in critical phenomena for the Einstein-Vlasov system and the
question of universality by explicitly exploiting the fact that
for this system the existence of an abundance of steady states is
well-established \cite{RR4}, and that these steady states can
easily be computed numerically. Computing a steady state $f_s$ we
consider the family $A f_s$ of initial data. Within every family
of steady states given by a specific dependence on particle energy
and angular momentum we find unstable ones which act as
critical solutions: If they are perturbed with $A>1$ they collapse
to a black hole, if they are perturbed with $A<1$ they either
disperse or oscillate in a neighborhood of the steady state,
depending on the sign of the binding energy. Due to the abundance
of possible such dependences on particle energy and angular
momentum there cannot be a universal critical solution in
spherically symmetric collapse for the Einstein-Vlasov system.

The paper proceeds as follows: In the next section we formulate
the Einstein-Vlasov system, first in general coordinates and then
in coordinates adapted to spherical symmetry. In \cite{RRS2}
Schwarzschild coordinates were used. Here we use maximal areal
coordinates, as was done in \cite{OC}. This has the advantage that
regions of spacetime containing trapped surfaces can be covered.
In Section~3 we discuss the numerical scheme which we use; it is a
particle-in-cell scheme of the type used for kinetic models in
plasma physics. It should be noted that for the analogous scheme
used in \cite{RRS2} a rigorous convergence proof has been
established in \cite{ReRo}, and we conjecture that this can be
done for the present scheme as well. In Section~4 we investigate
the stability properties of certain steady states and compare our
findings with the earlier work mentioned above. Section~5 is
devoted to critical phenomena and non-universality, and in a final
section we discuss the reliability of our code.

Our main motivation for this numerical analysis is that it may
lead to conjectures on the behavior of solutions of the Einstein-Vlasov
system which may eventually be proven rigorously.

\section{The Einstein-Vlasov system}
We first write down the Einstein-Vlasov system
in general coordinates on the tangent bundle $TM$ of the spacetime
manifold $M$.
Following standard practice
we normalize the physical constants to one.
The system then reads as follows:
\[
p^a \partial_{x^a} f - \Gamma^a_{bc}
p^b p^c  \partial_{p^a} f = 0,
\]
\[
G^{a b} = 8 \pi T^{a b},
\]
\[
T^{a b}
= \int p^a p^b f \,|g|^{1/2} \,\frac{d^4 p}{m} .
\]
Here $f$ is the number density of the particles on phase space,
$\Gamma^a_{b c}$ and $G^{a b}$
denote the Christoffel symbols and the Einstein tensor
obtained from the spacetime metric $g_{a b}$,
$|g|$ denotes its determinant,
$T^{a b}$ is the energy-momentum tensor generated by $f$,
$x^a$ are coordinates on $M$,
$(x^a,p^b)$ the corresponding coordinates on the tangent
bundle $TM$, Latin indices run from 0 to 3, and
\[
m=|g_{a b} p^a p^b |^{1/2}
\]
is the rest mass of a particle at the corresponding point in phase space.
We assume that all particles have rest mass 1 and move
forward in time so that the distribution function $f$ lives
on the mass shell
\[
PM = \Bigl\{ g_{a b} p^a p^b = -1,\ p^0 > 0 \Bigr\}.
\]
We consider this system in the spherically
symmetric, asymptotically flat case and write the metric
in the following form:
\[
ds^{2}=-(\alpha^2+a^2\beta^2)dt^2+2a^2\beta dtdr+a^2 dr^2
+ r^2\left(d\theta^2 + \sin^2\theta\, d\phi^2\right).
\]
Here the metric coefficients $\alpha, \beta$, and $a$ depend on
$t\in \R$ and $r\geq 0$,
$\alpha$ and $a$ are positive, and the polar angles $\theta\in[0,\pi]$
and $\phi\in[0,2\pi]$ parameterize the unit sphere.
The radial coordinate $r$ is thus the area radius.
Let $K^{a}_{b}$ be the second fundamental form and define
\[
\kappa:=K^{\theta}_{\theta}=\frac{\beta}{r\alpha}.
\]
By imposing the maximal gauge condition, which means that
each hypersurface of
constant $t$ has vanishing mean curvature, we obtain
the following field equations:
\begin{eqnarray}
&\displaystyle a_{r}=\frac{3}{2}a^{3}r \kappa^2+4\pi
r a^{3}\rho+\frac{a}{2r}(1-a^{2}),&\label{ee1}\\
&\displaystyle \kappa_{r}=-\frac{3}{r}\kappa-4\pi a\jmath,&\label{ee2}\\
&\displaystyle a_{t}=2\alpha a \kappa+(a\beta)_{r},&\label{ee3}\\
&\displaystyle
\alpha_{rr}=\alpha_{r}\left(\frac{a_{r}}{a}-\frac{2}{r}\right)+
\frac{2\alpha}{r^2}\left(2r\frac{a_{r}}{a}+a^2-1\right)+4\pi
a^2\alpha(S-3\rho).&\label{ee4}
\end{eqnarray}
The Vlasov equation takes the form
\begin{equation} \label{vlasov}
\partial_{t}f+\left(\frac{\alpha w}{a \epsilon}-\beta\right)\partial_{r}f
+\left(-\frac{\alpha_{r}  \epsilon}{a}-2\alpha \kappa w+ \frac{\alpha
L}{a r^3  \epsilon}\right)\partial_{w}f=0,
\end{equation}
where
\begin{equation}
 \epsilon= \epsilon(r,w,L)=\sqrt{1+w^{2}+L/r^{2}}.\label{E}
\end{equation}
The variables $w$ and $L$ can be thought of as the momentum in the
radial direction and the square of the angular momentum
respectively, see \cite{R} for more details. The matter
quantities are defined by
\begin{eqnarray}
\rho(t,r)
&=&
\frac{\pi}{r^{2}}
\int_{-\infty}^{\infty}\int_{0}^{\infty} \epsilon f(t,r,w,L)
\,dL\,dw, \label{rho}\\
\jmath(t,r)
&=&
\frac{\pi}{r^{2}}
\int_{-\infty}^{\infty}\int_{0}^{\infty}w f(t,r,w,L)\,dL\,dw,\label{j}\\
S(t,r)
&=&
\frac{\pi}{r^{2}}\int_{-\infty}^{\infty}\int_{0}^{\infty}
\frac{w^2 + L /r^{2}}{\epsilon}\, f(t,r,w,L)\,dL\,dw.\label{q}
\end{eqnarray}
We impose the following boundary conditions which ensure asymptotic flatness
and a regular center:
\begin{equation}
a(t,0)=a(t,\infty)=\alpha(t,\infty)=1.\label{bdryc}
\end{equation}
The equations (\ref{ee1})--(\ref{q}) together with the boundary conditions
(\ref{bdryc}) constitute the Einstein-Vlasov system for a spherically
symmetric, asymptotically flat spacetime
in maximal areal coordinates.

Let us consider some simple properties
of this system, in particular such as will be relevant for
its numerical simulation; a more careful mathematical analysis
of the system will be performed elsewhere \cite{AR1}.
First we note that the phase space density $f$ is constant along
solutions of the characteristic system
\begin{eqnarray}
\dot r
& = &
\frac{\alpha(\tau,r) w}{a(\tau,r)  \epsilon}-\beta(\tau,r),\label{charsys1}\\
\dot w
& = &
-\frac{\alpha_{r}(\tau,r)  \epsilon}{a(\tau,r)}-
2\alpha(\tau,r) \kappa(\tau,r) w +
\frac{\alpha(\tau,r) L}{a(\tau,r) r^3  \epsilon}, \label{charsys2}\\
\dot L
& = &
0 \label{charsys3}
\end{eqnarray}
of the Vlasov equation. If $\tau\mapsto (R,W,L)(\tau,t,r,w,L) =: Z(\tau,t,z)$
denotes the solution of the characteristic system with
$(R,W,L)(t,t,r,w,L)=(r,w,L)$ then
\[
f(t,r,w,L)=\fn((R,W,L)(0,t,r,w,L)),
\]
with $\fn = f(0,\cdot)$ the initial datum for $f$.
Due to our choice of coordinates $w,L$ in momentum space
the characteristic flow is not measure preserving. Indeed, if by $D$ we
denote differentiation along a characteristic of the Vlasov equation,
then
\begin{equation} \label{volelev}
D (f\,dL\,dw\,dr) = - \left(\frac{\alpha w}{a^2  \epsilon}a_r + \beta_r +
\frac{2 \beta}{r} \right)\,f\,dL\,dw\,dr;
\end{equation}
the factor on the right hand side is the $(r,w,L)$-divergence of the
right hand side of the characteristic system. Notice that by
Eqn.~(\ref{ee3}) this factor equals $-D \ln a$.
If $A\subset \R_+ \times \R \times \R_+$ is measurable and
$A(t):= Z(t,0,A)=\{(R,W,L)(t,0,r,w,L) \mid (r,w,L) \in A\}$
then
\[
\int_{A(t)} f(t,z)\, dz = \int_A \fn (z)\frac{a(0,r)}{a(t,R(t,0,z))} dz,
\]
while
\[
\int_{A(t)} a(t,r)\, f(t,z)\, dz
= \int_A a(0,r)\,\fn (z)\, dz.
\]
In particular, the total number
of particles, which, since all particles
have rest mass one, equals the rest mass of the system, is conserved:
\[
4\pi^2\int_{0}^{\infty}\int_{-\infty}^{\infty}\int_{0}^{\infty}
a(t,r)f(t,r,w,L)\,dL\,dw\,dr = M_{0}.
\]
Next we notice that Eqn.~(\ref{ee2}) can be rewritten as
\begin{equation} \label{ee2rewr}
\left(r^3 \kappa\right)_r = - 4 \pi r^3 a \jmath.
\end{equation}
Using Eqn.~(\ref{ee1}) the second order equation (\ref{ee4}) can be
rewritten as
\begin{equation} \label{ee4rewr}
\left(\frac{r^2}{a}\alpha_r \right)_r =
4 \pi r^2 a \alpha (\rho + S) + 6 r^2 a \alpha \kappa^2.
\end{equation}
The Hawking mass $m$ is given by
\begin{equation}  \label{m}
m=\frac{r}{2}\left(1+\frac{\beta^2}{\alpha^2}-\frac{1}{a^2}\right).
\end{equation}
We also introduce the quantity
\[
\mu:=\frac{r}{2}\left(1-\frac{1}{a^2}\right),
\]
and note that by Eqn.~(\ref{ee1}), $\mu$ can be written in the form
\[
\mu(t,r) =
\int_{0}^{r}\left(4\pi \rho(t,s)+\frac{3}{2}\kappa^2(t,s)\right) s^2 ds.
\]
Assuming that the matter is compactly supported
initially and hence also for later times
Eqn.~(\ref{ee2rewr}) implies that $\kappa(t,r)\sim r^{-3}$ for $r$ large.
Hence the limits as $r$ tends to $\infty$ of $m$ and $\mu$ are
equal so that the ADM mass
$M$ can be written as
\[
M=\int_{0}^{\infty}\left(4\pi \rho(t,r)+\frac{3}{2}
\kappa^2(t,r)\right) r^2 dr.
\]
The ADM mass is conserved.

It is of interest to find out when trapped surfaces form, which means
that $2m/r>1,$ and in view of (\ref{m}) this
condition can be written as
\[
a^2>\frac{\alpha^2}{\beta^2}.
\]
Note also that $2m/r<1/2$ implies that $2\mu/r<1/2$. From \cite{MO} we
obtain the following inequalities
\begin{equation} \label{os}
\left|\frac{1}{a}-\frac{\beta}{\alpha}\right|\leq 1,\;\;
\left|\frac{1}{a}+\frac{\beta}{\alpha}\right|\leq 1.
\end{equation}
Finally we notice that
the second order equation for $\alpha$ in the form (\ref{ee4rewr})
can be integrated:
\begin{equation} \label{alphar}
\alpha_{r}(t,r)
= \frac{a(t,r)}{r^2}\int_{0}^{r}\left(4\pi\alpha a (\rho+S)+6 a \alpha
\kappa^2\right) s^2 ds.
\end{equation}
Thus $\alpha$ is monotonically increasing outwards, and from (\ref{bdryc})
and (\ref{os}) it follows that
\[
0<\alpha\leq 1,\ |\beta|\leq\alpha\leq 1.
\]

\section{The numerical method}
Let us consider an initial condition $\fn=\fn(r,w,L)$, define
additional variables $u\geq 0$ and $\phi \in [0,\pi]$
by the relations
\[
u^2 = w^2 + \frac{L}{r^2},\ w = u\, \cos \phi,\ L = r^2 u^2 \sin^2 \phi
\]
and assume that $\fn$ vanishes outside the set
$(r,u, \phi ) \in [R_0,R_1] \times [U_0,U_1] \times
[ \Phi_0, \Phi_1]$. We will approximate the solution
using a particle method.  For a thorough treatment of particle methods
in the context of plasma physics we refer to \cite{bl}.
To initialize the particles we take integers $N_r,N_u, N_{\phi}$ and define
\[
\Delta r =
\frac{R_1 - R_0}{N_r},\
\Delta u = \frac{U_1 - U_0}{N_u},\
\Delta \phi = \frac{\Phi_1 - \Phi_0}{N_{\phi}},
\]
\[
r_i =
R_0 + \left(i-\frac{1}{2}\right) \Delta r, \ u_j = U_0+
\left(j-\frac{1}{2}\right) \Delta u,\ \phi_k = \Phi_0 +
\left(k-\frac{1}{2}\right)\Delta \phi,
\]
\[
f^0_{ijk} =
\fn\left(r_i, u_j, \phi_k \right)\; 4 \pi r^2_i
\Delta r\; 2\pi u^2_j \Delta u\, \sin \phi_k \Delta\phi,
\]
\[
r^0_{ijk} =
r_i,\ w^0_{ijk} = u_j\,\cos \phi_k,\ L_{ijk} =
\left( r_i u_j\,\sin \phi_k \right)^2.
\]
At this point it should be noted that $f^0_{ijk}$
contains the phase space volume element, a fact which is convenient
when computing the induced components of the energy momentum tensor,
but which has to be observed when this quantity is time-stepped.
At each point in phase space with coordinates
$r^0_{ijk},w^0_{ijk},L_{ijk}$ we imagine a particle with
weight $f^0_{ijk}$ which is smeared out in the radial direction,
i.e., with respect to $r$ it is represented by a hat function
of width $2 \Delta r$. Once the particles are initialized
the grid covering the support of the initial datum plays no further
role.
From these numerical particles,
approximations are made of the quantities $\rho,\ \jmath,\ S$
defined in (\ref{rho}), (\ref{j}), (\ref{q})
at the spatial grid points
\[
r_j := j \Delta r,\ j=0,\ldots, N+1;
\]
note that this is now a new grid, which we extend from $r=0$ to
the radius of the spatial support of the matter quantities.
The field equations (\ref{ee1}) and (\ref{ee2rewr}),
which do not contain the metric coefficient $\alpha$,
can now be integrated on the support of the matter
from the origin outward, using
the boundary conditions (\ref{bdryc}) and $(r^3 \kappa)_{|r=0}=0$;
notice that $\kappa$ is then also known outside the support of the
matter.

To obtain an approximation of the metric coefficient $\alpha$
at the grid points  $r_j$ we discretize Eqn.~(\ref{ee4rewr})
in the following way: We write
\[
r_{j+\frac{1}{2}} := \left(j+\frac{1}{2}\right) \Delta r,
\]
and from what is already computed we obtain approximations
for $a$, $\kappa$, $\rho$, $S$ both at the grid points $r_j$
and by interpolation at
the intermediate points $r_{j+\frac{1}{2}}$, which
we denote by
\[
a_j,\ a_{j+\frac{1}{2}},\ \ldots
\]
Using the approximations
\[
\left(\frac{r^2}{a}\alpha_r\right)_r (r_j)
\approx
\frac{1}{\Delta r} \left[
\frac{r_{j+\frac{1}{2}}^2 \alpha_r(r_{j+\frac{1}{2}})}{a_{j+\frac{1}{2}}}
-
\frac{r_{j-\frac{1}{2}}^2 \alpha_r(r_{j-\frac{1}{2}})}{a_{j-\frac{1}{2}}}
\right]
\]
and
\[
\alpha_r(r_{j+\frac{1}{2}}) \approx 
\frac{\alpha_{j+1} - \alpha_{j}}{\Delta r},\
\alpha_r(r_{j-\frac{1}{2}}) \approx 
\frac{\alpha_{j} - \alpha_{j-1}}{\Delta r}
\]
we obtain the following linear system for the approximations
$\alpha_{j}$ for $\alpha(r_j)$ on the grid points:
For $j=0,\ldots, N-1$,
\begin{eqnarray*}
\frac{r_{j+\frac{1}{2}}^2}{a_{j+\frac{1}{2}}} \alpha_{j+1}
&+& \frac{r_{j-\frac{1}{2}}^2}{a_{j-\frac{1}{2}}} \alpha_{j-1} -
\\
&-& \left[\frac{r_{j+\frac{1}{2}}^2}{a_{j+\frac{1}{2}}} +
\frac{r_{j-\frac{1}{2}}^2}{a_{j-\frac{1}{2}}}
+ \left[6 \kappa_j^2 + 4 \pi(\rho_j + S_j)\right]r_j^2 (\Delta r)^2 a_j \right]
\alpha_j
= 0,
\end{eqnarray*}
and
\[
\alpha_1 - \alpha_0 = 0,\
\alpha_{N+1} = \sqrt{1-\frac{2 M^\ast}{r_{N+1}}}.
\]
The last two equations arise from the boundary condition
\[
\left(\frac{r^2}{a}\alpha_r\right)_r (t,0) = 0
\]
and the approximation
\[
\alpha(t,r) \approx \sqrt{1-\frac{2 \mu(t,r)}{r}}
\]
for $r$ large, in particular, $r$ well outside the
support of the matter. The latter approximation is motivated by the
known representation of the asymptotically flat Schwarzschild
solution in maximal areal coordinates. The approximation $M^\ast$ for
$\mu(t,r)$ at large values of $r$ can be computed
from the already approximated quantities $\rho$ and $\kappa$.

The linear system for $\alpha_j$ is tridiagonal, obviously
diagonally dominated, and it can easily be solved.
Thus we now have approximations for all the field quantities
on the spatial grid points $r_j$.
In passing we note that in \cite{OC} the field equation
(\ref{ee4}) was discretized directly resulting again in a
tridiagonal system for $\alpha_j$, but our approach of
discretizing the rewritten version (\ref{ee4rewr}) instead
seems to perform better numerically.

To perform the time step we propagate the numerical particles
according to the characteristic system of the Vlasov
equation (\ref{charsys1}), (\ref{charsys2}), (\ref{charsys3}). To do so we
interpolate the field quantities to particle
locations and use a simple Euler time stepping method
to define the new
phase space coordinates $r^1_{ijk}, w^1_{ijk}, L^1_{ijk}$
of the numerical particle with label $ijk$.
We still need to propagate the phase space volume element
along the characteristics.
Discretizing (\ref{volelev}) in time we obtain the relation
\[
\frac{1}{\Delta t} \left(f^1_{ijk} - f^0_{ijk} \right) = -
\left(\frac{\alpha w}{a^2  \epsilon}a_r + \beta_r +
\frac{2 \beta}{r} \right)\, f^0_{ijk}
\]
which we use to compute $f^1_{ijk}$, and one time step is complete.

\section{Stability issues for steady states}
For static solutions, $\beta = \kappa = \jmath = 0$ so that
\[
m(r) = \mu(r) = 4 \pi \int_0^r \rho(s)\, s^2 ds,
\]
and the field equation (\ref{ee1}) decouples and is
solved by
\begin{equation} \label{astat}
a(r) = \left(1-\frac{2 m(r)}{r}\right)^{-1/2}.
\end{equation}
The metric coefficient $\alpha$ is determined by the
equation
\begin{equation} \label{alphastat}
\frac{\alpha'}{\alpha} = a^2 \left(\frac{m}{r^2} + 4 \pi r p\right)
\end{equation}
where $\alpha' = \alpha_r = d\alpha/dr$ and 
\[
p(r) :=
\frac{\pi}{r^{2}}\int_{-\infty}^{\infty}\int_{0}^{\infty}
\frac{w^2}{ \epsilon} f(r,w,L)\,dL\,dw
\]
is the radial pressure. Eqn.~(\ref{alphastat}) is the $rr$-component
of the Einstein equations, and a lengthy computation using the
Tolman-Oppenheimer-Volkov equation
\[
p' = - \frac{\alpha'}{\alpha} (p + \rho) - \frac{2}{r}(p-p_T)
\]
shows that the second order field equation (\ref{ee4}) follows;
\[
p_T(r) := \frac{\pi}{2 r^{2}}\int_{-\infty}^{\infty}\int_{0}^{\infty}
\frac{L}{ \epsilon r^2} f(r,w,L)\,dL\,dw
\]
is the tangential pressure. It is easy to check that in addition
to $L$ also the particle energy
\[
E:=\alpha(r) \sqrt{1+w^{2}+\frac{L}{r^2}} = \alpha(r) \epsilon
\]
is constant along characteristics in the static case.
Hence the ansatz
\begin{equation}
f(r,w,L)=\Phi(E,L), \label{ansats}
\end{equation}
satisfies the static Vlasov equation. By substituting it into
the definitions for $\rho$ and $p$ these quantities become
functionals of $\alpha$, and the static Einstein-Vlasov system
is reduced to the equation (\ref{alphastat}) with
(\ref{astat}) substituted in. Given some ansatz function
$\Phi$ and prescribing some value for $\alpha(0)$,
the static system can therefore
easily be solved numerically, by integrating (\ref{alphastat})
from $r=0$ outward,
in particular if $\Phi$ is such that the resulting dependence
of $\rho$ and $p$ on $\alpha$ can be computed explicitly.
This is the case for the power law
\begin{equation}
f(r,w,L)=\Phi(E,L)=(E_0-E)^k_{+}(L-L_0)_{+}^l.\label{pol}
\end{equation}
Here $l>-1/2,\,k>-1,\,L_0\geq 0$, $E_0>0$ is the cut-off energy,
and $x_{+}:=\max\{x,0\}$.
In the Newtonian case with $l=L_0=0$ this ansatz leads
to steady states with a polytropic equation of state.
Note that by taking $L_{0}>0$ there will be no matter in the
region
\[
r<\sqrt{\frac{L_0}{(E_{0}/\alpha(0))^2-1}},
\]
since there necessarily
$E>E_{0}$ and $f$ vanishes. We call such configurations static
shells. The existence of static shells with finite mass and finite
extension has been proved in \cite{R1}. It is numerically
convenient to work with shells since potential difficulties in
treating $r=0$ are avoided, and we will only consider shells here.
This is also motivated by the fact that in the numerical
experiments performed on critical phenomena in \cite{RRS2}
the initial data for the matter had such a shell structure
so that static shells are natural candidates for the critical
solutions.
It should be pointed out that there are static solutions which do not
have the form (\ref{ansats}), cf.\ \cite{Sc2}.

In our simulations we have studied four cases:\smallskip \newline
\hspace*{1cm}
\begin{tabular}{ll}
Case 1:& $k=0,\, l=0$,\\
Case 2:& $k=0,\, l=1/2$,\\
Case 3:& $k=1,\, l=1/2$,\\
Case 4:& $k=0,\, l=3/2$.
\end{tabular} \smallskip\\
Having chosen $k$ and $l$ we then
numerically construct static solutions to the Einstein-Vlasov
system as indicated above,
by specifying values on $E_{0},\, L_{0}$ and $\alpha(0)$.
The resulting metric coefficient will a-priori not
satisfy the boundary condition $\alpha(\infty) = 1$,
but by shifting both $\alpha$ and $E_0$ appropriately
a steady state which satisfies the boundary conditions
is obtained.
The distribution function $f_s$ of the steady state
is then multiplied by an amplitude $A,$ so that a new, perturbed
distribution function is obtained. This is then used as initial
datum in our evolution code. Accordingly, if we choose $A=1$ then
the initial datum is exactly the steady state, and a good test of
our code is to check how much it deviates from being static
in the evolution.
We find that such initial data are tracked extremely well which is
a very satisfying feature of our code.
Of course, for unstable steady
states the numerical errors introduced will make the solution
drift off after some time which can be made longer by increasing
the number of numerical particles and the number of time steps.

For $k$ and $l$ fixed we characterize
each steady state by its
central red shift $Z_{c}$ and its fractional binding energy
$E_{b},$ which are defined by
\[
Z_{c}=\frac{1}{\sqrt{\alpha(0)}}-1,\;\; E_{b}=\frac{e_b}{M_{0}},
\mbox{ where }\;\; e_{b}=M_{0}-M.
\]
The central redshift is the redshift of a photon emitted from the
center and received at infinity, and the binding energy $e_b$ is
the difference of the rest mass and the ADM mass and is a
conserved quantity. In Figure~1 and Figure~2 below the relation between
the fractional binding energy and the central redshift is given
for Case 1 and Case 4 with $L_0=0.1$.
\begin{figure}[htbp]
\begin{center}\scalebox{.6}{\includegraphics{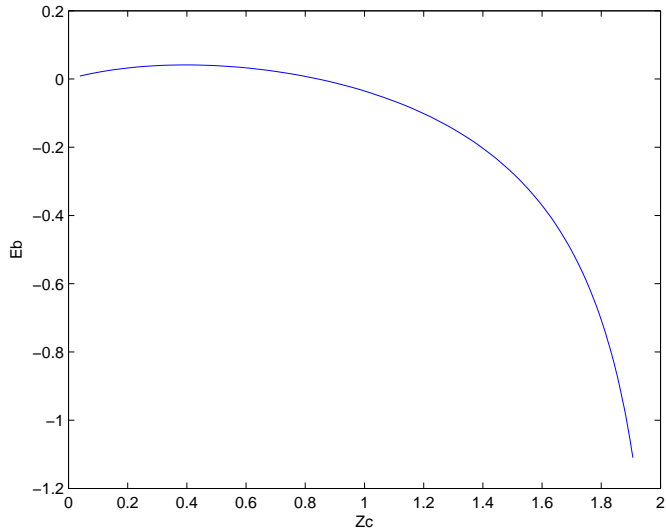}}
\end{center}
\caption{Case 1 with $L_0=0.1$}\label{fig1}
\end{figure}

\begin{figure}[htbp]
\begin{center}\scalebox{.6}{\includegraphics{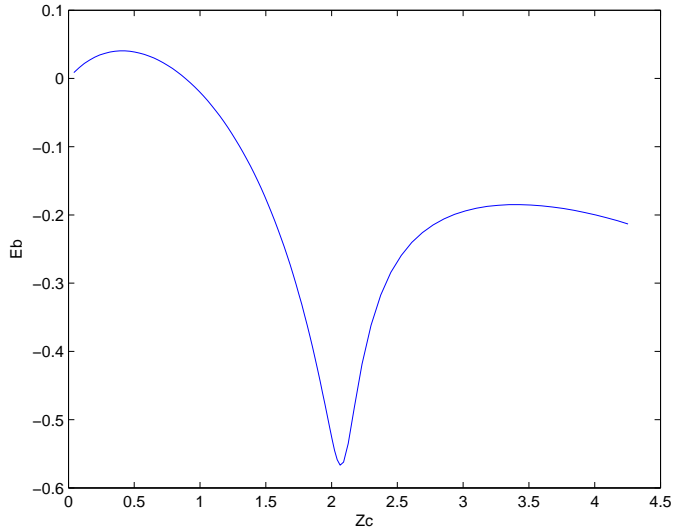}}
\end{center}
\caption{Case 4 with $L_0=0.1$}\label{fig2}
\end{figure}

The relevance of these concepts for the stability properties of
steady states was first discussed by Zeldovich and Podurets
\cite{ZP} who argued that it should be
possible to diagnose the stability from binding energy
considerations. Zeldovich and Novikov \cite{ZN} then conjectured
that the binding energy maximum
along a steady state sequence signals the onset of instability.
Ipser \cite{I1} and Shapiro and Teukolsky \cite{Sh2} find
numerical support for this conjecture and they also find that
steady states with a central redshift above about $0.5$ result in
collapse and are thus unstable. In both of these studies only
isotropic steady states were considered, i.e. $\Phi=\Phi(E),$
whereas our study includes the dependence on $L.$ Moreover, our
algorithm is closely related to the one in \cite{RRS2} for which
convergence has been proved in \cite{ReRo}.
Of course, we also take advantage of
the great progress of the computer capacity in that we can use a
larger number of particles and considerably improve the resolution
of the grid compared to what was possible in the earlier simulations. To get
an indication of the accuracy of our simulation we compute at
every time step the ADM mass $M$ and the rest mass $M_0$, both of
which should be conserved, and \textit{as long as no trapped
surface has formed the errors are remarkably small}. We will get
back to a discussion of the numerical errors in the final section.

Before describing the results of our simulations we mention that
besides perturbations of a steady state $f_s$ of the form $A f_s$
we also considered perturbations of the form
$f_s(r+r_{sh},w,L)$ and $f_s(r,w+w_{sh},L)$ where the state
is shifted with respect to $r$ by $r_{sh}$ or with respect
to $w$ by $w_{sh}$. It turns out that perturbations with
$A>1$ or $r_{sh}<0$ or $w_{sh}<0$, which in principle push
the state towards collapse, and on the other hand 
perturbations with
$A<1$ or $r_{sh}>0$ or $w_{sh}>0$, which in principle push
the state towards dispersion, lead to the same qualitative
features of the perturbed solution. In the following discussion
we restrict ourselves to perturbations of the form $A f_s$.

The general picture that arises from our simulations is summarized
in Tables~1--4 which correspond to the four different cases we
consider. The parameter $L_0$ is the same in all cases, i.e.
$L_{0}=0.1.$
\begin{table}[h]
\centering
\begin{tabular}{|r|r|c|c|} \hline
$Z_c$ & $E_b$ & $A<1$ & $A>1$ \\ \hline
$0.24$  & $0.036$ & stable & stable \\
$0.30$  & $0.039$ & stable & stable \\
$0.39$  & $0.041$ & stable & stable \\
$0.43$  & $0.041$ & stable & unstable \\
$0.47$  & $0.040$ & stable & unstable \\
$0.52$  & $0.038$ & stable & unstable \\
$0.65$ & $0.027$ & stable & unstable \\
$0.82$  & $0.004$ & stable & unstable \\
$0.95$  & $-0.024$ & unstable & unstable \\
$1.1$ & $-0.070$ & unstable & unstable \\ \hline
\end{tabular}
\caption{Case 1: $k=0$ and $l=0.$}
\end{table}

\begin{table}[h]
\centering
\begin{tabular}{|r|r|c|c|} \hline
$Z_c$ & $E_b$ & $A<1$ & $A>1$ \\ \hline
$0.21$  & $0.032$ & stable & stable \\
$0.34$  & $0.040$ & stable & stable \\
$0.39$  & $0.040$ & stable & stable \\
$0.42$  & $0.041$ & stable & unstable \\
$0.46$  & $0.040$ & stable & unstable \\
$0.56$  & $0.036$ & stable & unstable \\
$0.65$  & $0.029$ & stable & unstable \\
$0.82$  & $0.008$ & stable & unstable \\
$0.95$  & $-0.015$ & unstable & unstable \\
$1.20$  & $-0.078$ & unstable & unstable \\ \hline
\end{tabular}
\caption{Case 2: $k=0$ and $l=1/2.$}
\end{table}

\begin{table}[h]
\centering
\begin{tabular}{|r|r|c|c|} \hline
$Z_c$ & $E_b$ & $A<1$ & $A>1$ \\ \hline
$0.21$  & $0.030$ & stable & stable \\
$0.33$  & $0.037$ & stable & stable \\
$0.38$ & $0.038$ & stable & stable \\
$0.42$  & $0.039$ & stable & stable \\
$0.045$  & $0.039$ & stable & unstable \\
$0.052$  & $0.037$ & stable & unstable \\
$0.77$  & $0.020$ & stable & unstable \\
$0.90$  & $0.004$ & stable & unstable \\
$1.0$ & $-0.01$ & unstable & unstable \\
$1.13$ & $-0.037$ & unstable & unstable \\ \hline
\end{tabular}
\caption{Case 3: $k=1$ and $l=1/2.$}
\end{table}

\begin{table}[h]
\centering
\begin{tabular}{|r|r|c|c|} \hline
$Z_c$ & $E_b$ & $A<1$ & $A>1$ \\ \hline
$0.21$  & $0.032$ & stable & stable \\
$0.34$  & $0.039$ & stable & stable \\
$0.41$  & $0.040$ & stable & stable \\
$0.44$  & $0.040$ & stable & stable \\
$0.52$  & $0.038$ & stable & unstable \\
$0.69$  & $0.025$ & stable & unstable \\
$0.80$  & $0.014$ & stable & unstable \\
$0.92$  & $-0.004$ & unstable & unstable \\
$1.1$  & $-0.045$ & unstable & unstable \\ \hline
\end{tabular}
\caption{Case 4: $k=0$ and $l=3/2.$}
\end{table}

If we first consider perturbations with $A>1$ we find
that steady states with small values on $Z_c$ (less than
approximately $0.40$ depending on which case we consider) are 
stable, i.e., the perturbed solutions
stay in a neighborhood of the static solution
as depicted in Figure~3.
\begin{figure}[htbp]
\begin{center}\scalebox{.9}{\includegraphics{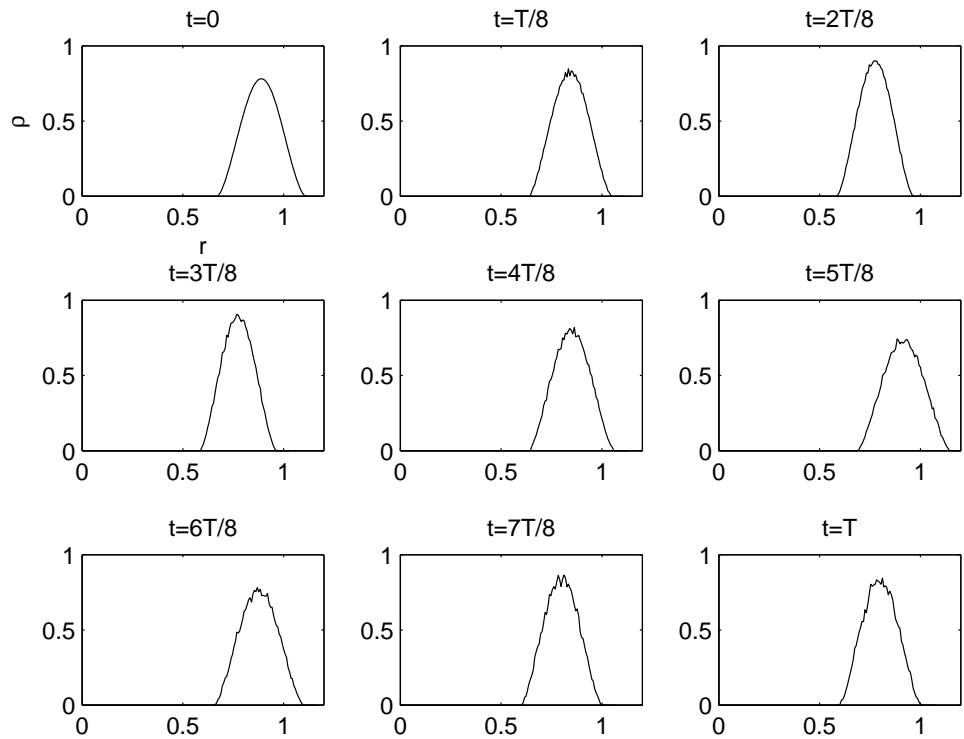}}
\end{center}
\caption{$Z_c=0.30,\ E_b=0.04,\ A=1.01,\ T=80.0$}\label{fig3}
\end{figure}
A more careful investigation of these perturbed solutions
seems to indicate that they oscillate in an (almost) periodic way,
and we come back to this issue at the end of this section.
For larger values of $Z_c$ the evolution leads to the formation
of trapped surfaces and by the result in \cite{DR} to the collapse to
black holes, as depicted in Figure~4. 
\begin{figure}[htbp]
\begin{center}\scalebox{.9}{\includegraphics{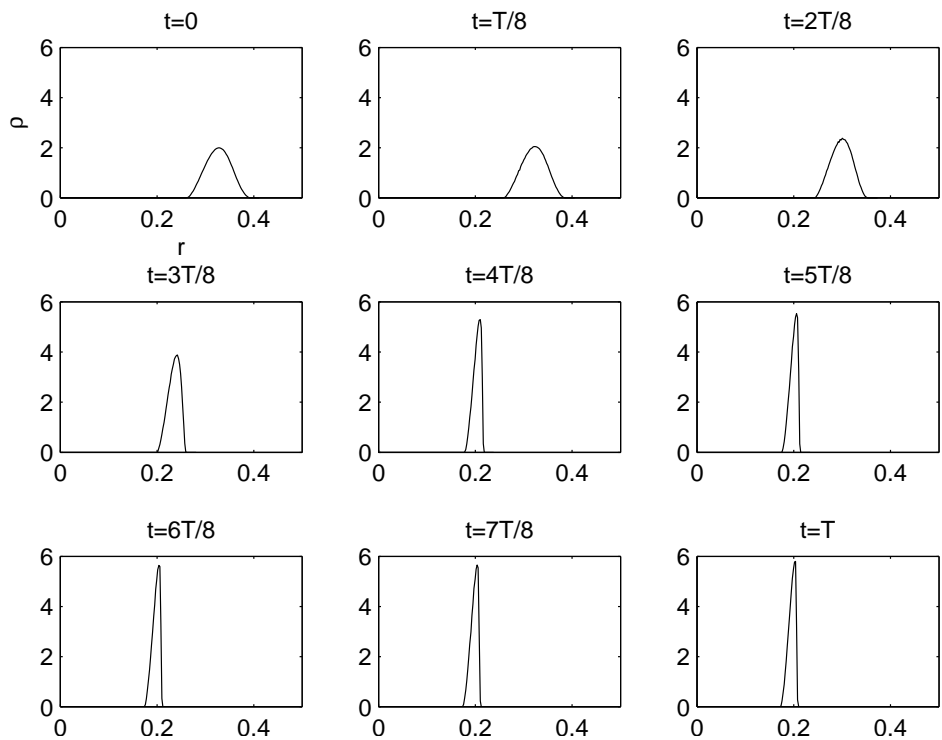}}
\end{center}
\caption{$Z_c=1.14,\ E_b=-0.08,\ A=1.01,\ T=10.0$}\label{fig4}
\end{figure}
As a matter of fact we
do check if null geodesics that start after a certain time from
the center are caught so that they cannot escape to infinity, and
our results always support that there is an event horizon.

Hence, for perturbations with $A>1$ the value of $Z_c$ alone seems
to determine the stability features of the steady states. By
plotting the curve, $E_b$ versus $Z_c,$ for a shorter interval of
$Z_c$ to get better resolution (Figures~5 and 6 below) we find
that the conjecture by Novikov and Zeldovich, that the maximum of
$E_b$ along a sequence of steady states signals the onset of
instability, is true in a numerical sense also for states that
depend on the angular momentum $L$.
\begin{figure}[htbp]
\begin{center}\scalebox{.6}{\includegraphics{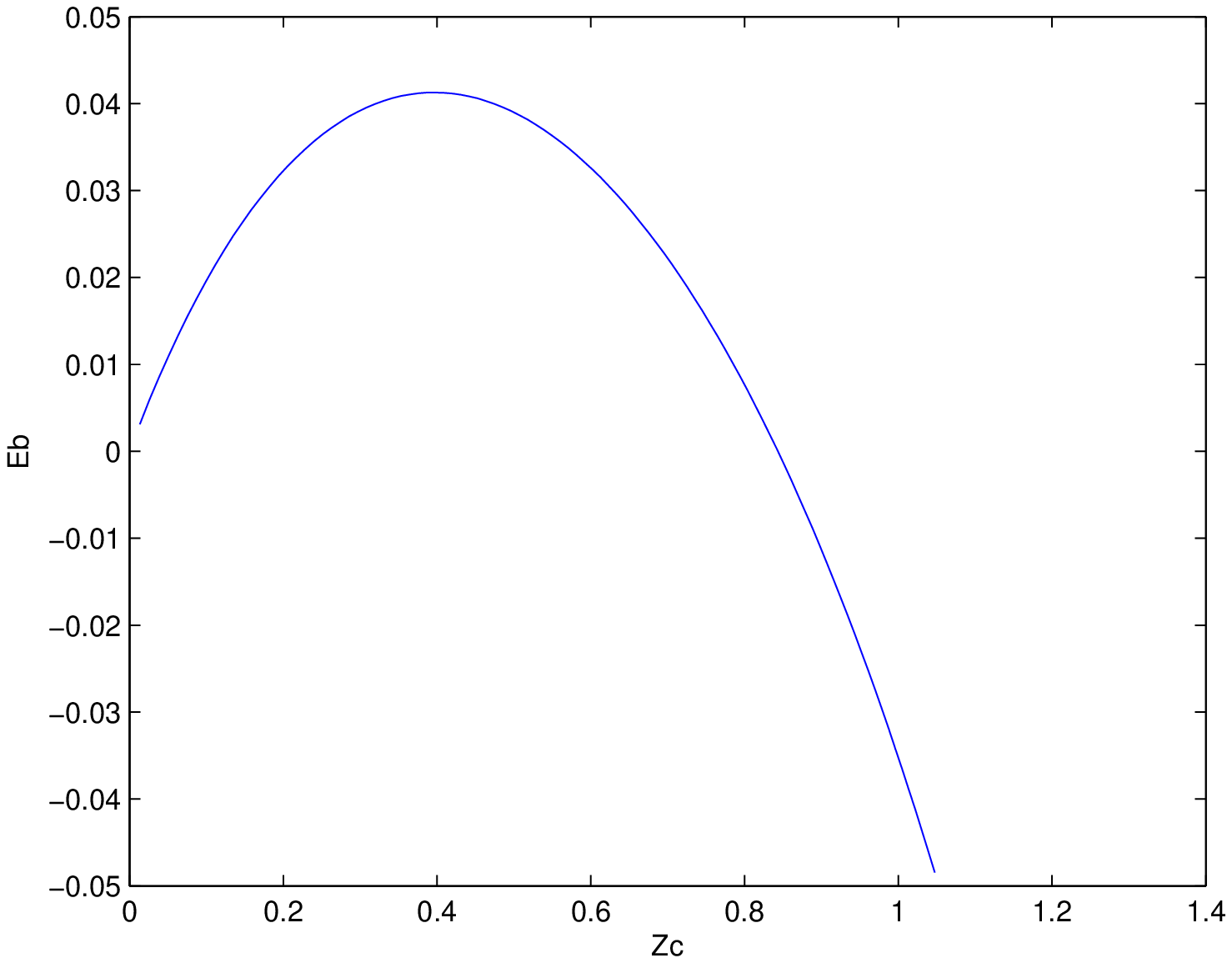}}
\end{center}
\caption{Case 1 with $L_0=0.1$}\label{fig5}
\end{figure}
\begin{figure}[htbp]
\begin{center}\scalebox{.6}{\includegraphics{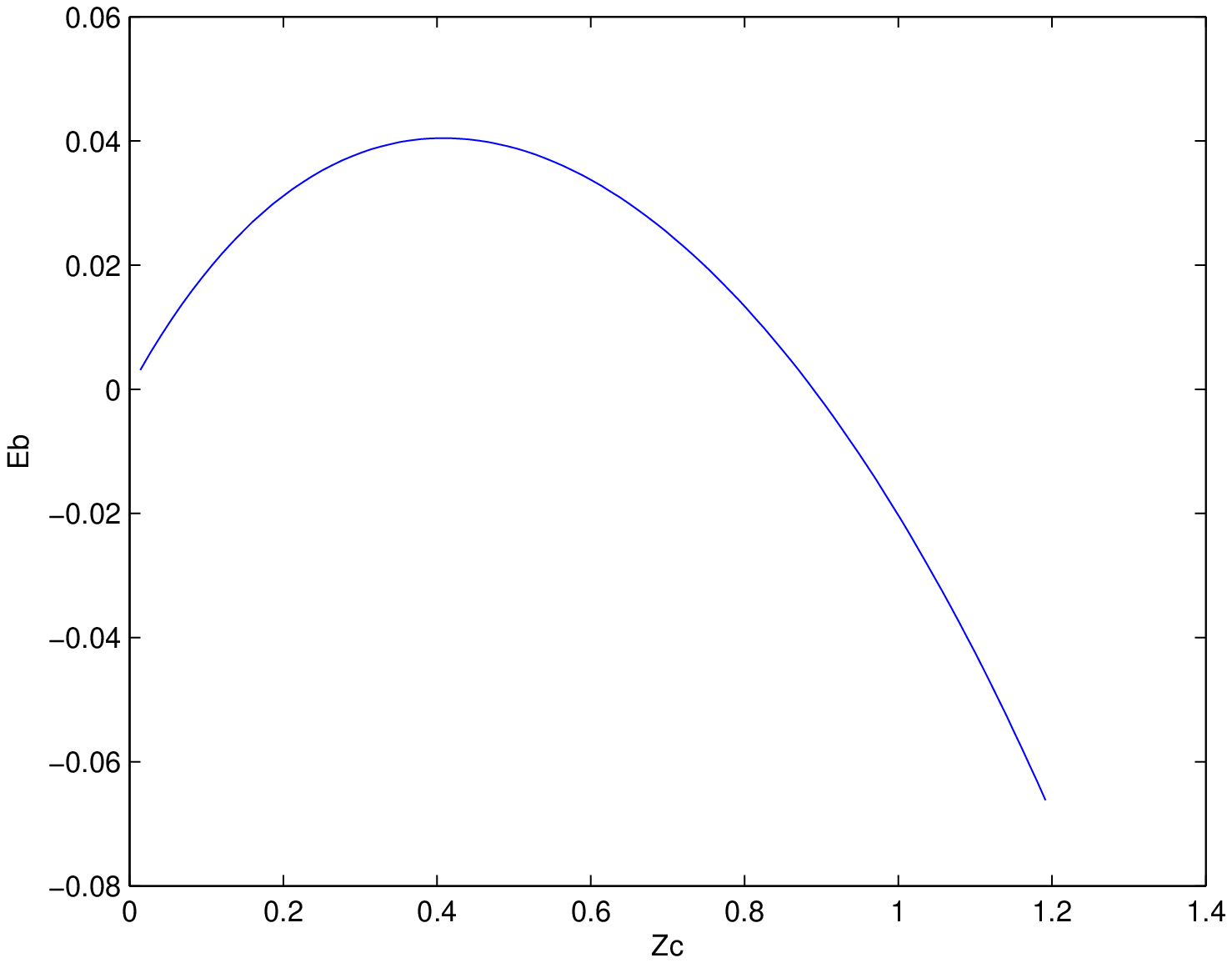}}
\end{center}
\caption{Case 4 with $L_0=0.1$}\label{fig6}
\end{figure}
Having a second look at Figure~2 one might think that
along that curve another change of stability behavior
might occur at $Z_c\approx 2$, but we found no indication
of this. The reason for the qualitative difference between
Figure~1 and Figure~2 and its possible consequences will be
investigated elsewhere \cite{AR2}.

The situation is quite different for perturbations with $A<1$.
The crucial quantity in this case is the fractional binding energy
$E_b$. Consider a steady state with $E_b >0$
and a perturbation with $A<1$ but close to $1$
so that the fractional binding energy remains positive. Then the
perturbed solution drifts
outwards and then turns back and reimplodes and comes close to its
initial state, and then continues to expand and reimplode and thus
oscillates. This is depicted in Figure~7, see also the end of this section.
\begin{figure}[htbp]
\begin{center}\scalebox{.9}{\includegraphics{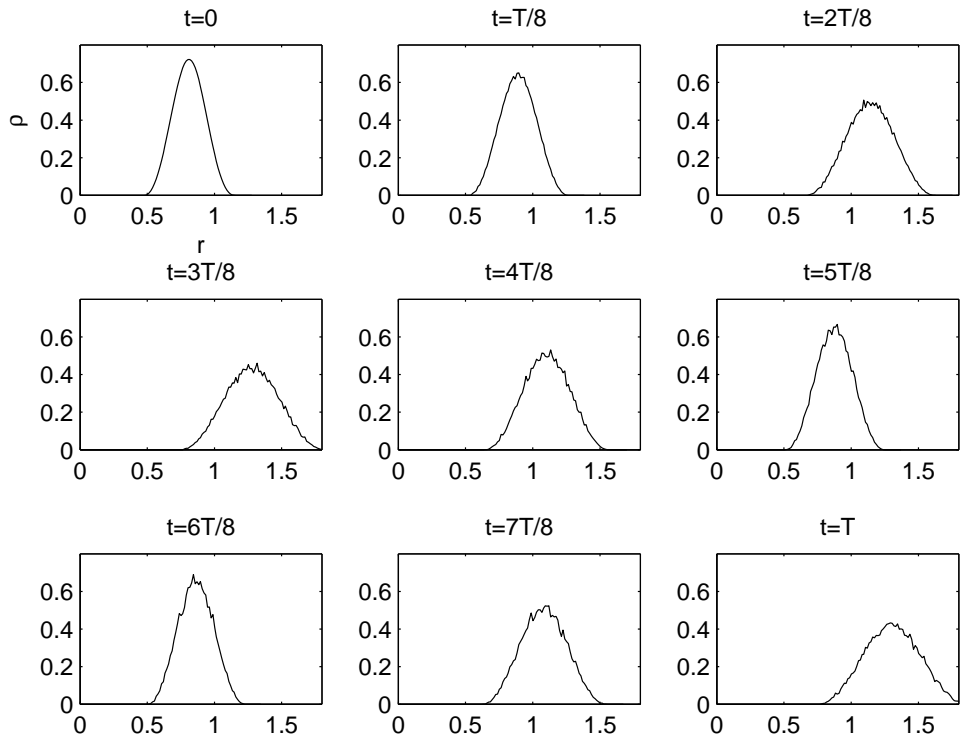}}
\end{center}
\caption{$Z_c=0.47,\ E_b=0.04,\ A=0.99,\ T=90.0$}\label{fig7}
\end{figure}

In \cite{Sh2} it is stated that if $E_b>0$ the solution must ultimately
reimplode. We are not aware of any precise mathematical
formulation (or proof) of such a statement but our simulations
support that it is true. For negative values of $E_b$ the
solutions with $A<1$ disperse to infinity as depicted in Figure~8.

\begin{figure}[htbp]
\begin{center}\scalebox{.9}{\includegraphics{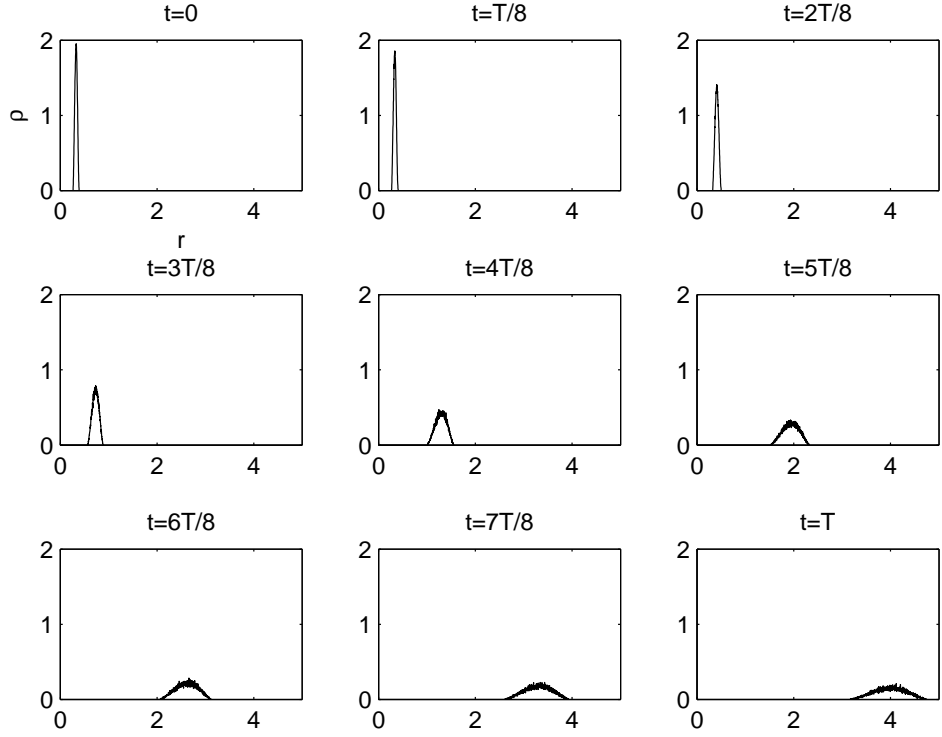}}
\end{center}
\caption{$Z_c=1.14,\ E_b=-0.08,\ A=0.99,\ T=14.0$}\label{fig8}
\end{figure}

A simple argument which relates the binding energy to the question
whether a solution disperses or not at least for the
case where the spatial support is a shell is the following:
Consider a solution which has an expanding vacuum region
of radius $R(t)$ at the center with $R(t) \to \infty$
for $t \to \infty$. Such a solution disperses in a strong
sense, and we claim that this implies that $M_0 \leq M$, i.e. $E_b \leq 0$.
To see this, observe that
\begin{eqnarray*}
M_0
&=&
4 \pi^2 \int_{R(t)}^\infty \int_{-\infty}^\infty \int_0^\infty
a(t,r)\, f(t,r,w,L)\, dL\,dw\,dr\\
&=&
4 \pi^2 \int_{R(t)}^\infty \left(1-\frac{2\mu(t,r)}{r}\right)^{-1/2}
\int_{-\infty}^\infty \int_0^\infty
f(t,r,w,L)\, dL\,dw\,dr\\
&\leq&
\left(1-\frac{2 M}{R(t)}\right)^{-1/2} 4 \pi^2 \int_{R(t)}^\infty
\int_{-\infty}^\infty \int_0^\infty
\epsilon f(t,r,w,L)\, dL\,dw\,dr\\
&=&
\left(1-\frac{2 M}{R(t)}\right)^{-1/2} M,
\end{eqnarray*}
so that with $t\to\infty$ necessarily $M_0 \leq M$ as claimed.

For the Vlasov-Poisson system, which is the Newtonian limit
of the Einstein-Vlasov system \cite{RR2} a rigorous stability
theory
for steady states of the form analogous to
(\ref{ansats}) has been developed in recent years,
based on variational techniques, cf.~\cite{GR1,GR2,GR3,GR4,R2}.
The result is that the stability properties of the steady state
are essentially determined by the ansatz function $\Phi$ in (\ref{ansats}),
in particular, all the polytropic steady states resulting from
an ansatz of the form (\ref{pol}) with $0<k<7/2,\ l=0,\ L_0=0$
are nonlinearly stable. The above discussion shows that an analogous
result does not hold for the Einstein-Vlasov system,
since there the stability depends in addition on the central
redshift and the binding energy, even if $k$ and $l$ are fixed.
We note at this point that the stability of shell like steady
states has not yet been systematically investigated for the Vlasov-Poisson
system. But we would argue that this does not affect the above
statement, since on the one hand it is reasonable to expect that
the stability results for shells in the Newtonian case will
be analogous to the case $L_0=0$, an expectation supported 
by recent results in this direction \cite{Schu}, 
and on the other hand
we do not expect our numerical stability results for the
relativistic case to depend on the shell structure of the
steady states. A clear indication that stability results
based on variational techniques do not carry over
to the relativistic case
without making additional restrictions
are the first results in this direction in \cite{Wo}.

\begin{figure}[htbp]
\begin{center}\scalebox{.6}{\includegraphics{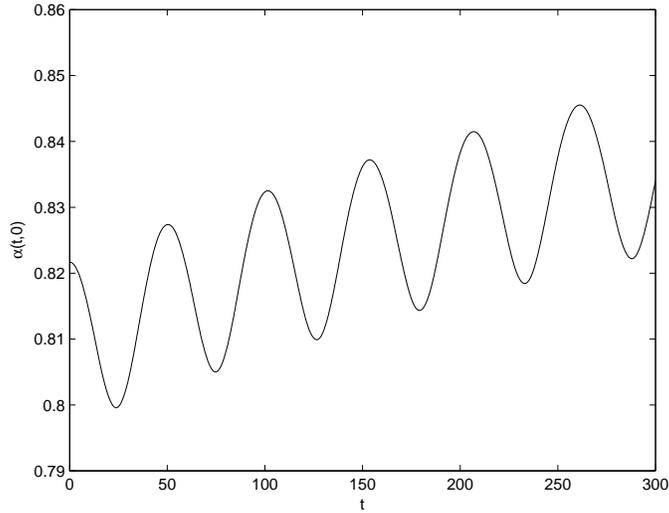}}
\end{center}
\caption{$Z_c=0.21,\ E_b=0.03,\ A=1.02,\ T=300.0$}\label{figperiodalpha}
\end{figure}

\begin{figure}[htbp]
\begin{center}\scalebox{.6}{\includegraphics{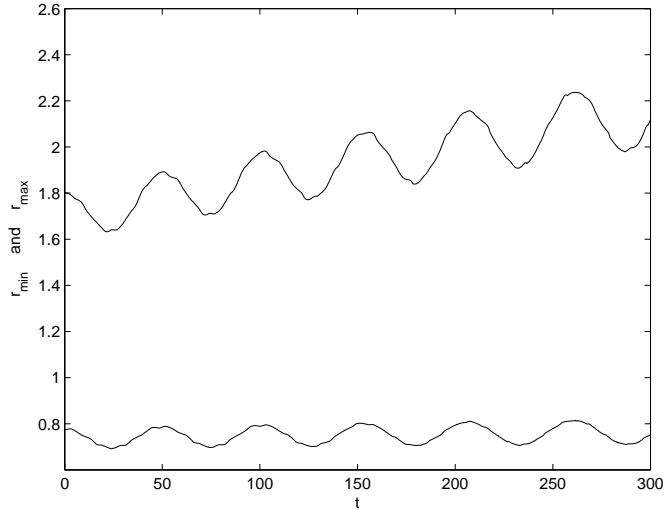}}
\end{center}
\caption{$Z_c=0.21,\ E_b=0.03,\ A=1.02,\ T=300.0$}\label{figperiodrminmax}
\end{figure}

Our investigation indicates that there are (at least)
three qualitatively different types of behavior
which can result from the perturbation of a steady state.
Perturbing an unstable state with $Z_c > 0.4$
using $A>1$ leads to gravitational collapse and a black hole.
Perturbing an unstable state with $E_b<0$
using $A<1$ leads to dispersion. In the other, stable cases
$Z_c < 0.4$ and $A>1$ or $E_b>0$ and $A<1$  
the perturbation leads to an oscillatory behavior. 
Whether the perturbed solution is indeed time-periodic
is hard to decide numerically. One way to
shed some light on this question is to plot 
$\alpha(t,0)$ or the inner and outer radius
$r_\mathrm{min}(t)$ and $r_\mathrm{max}(t)$ of the
matter support. The results are shown in 
Figures~\ref{figperiodalpha} and~\ref{figperiodrminmax}
for a steady state with $k=0, l=1/2$ (Case~2)
and $L_0=0.1,\ Z_c=0.21,\ E_b=0.03$, perturbed with $A=1.02$.
The oscillatory behavior with a clear time period
is quite striking. 
We found that all the solutions arising by a small perturbation of a
fixed steady state seem to have roughly the same period,
independently of $A$. We do at this point venture no interpretation
of the slow upward drift of $\alpha(t,0)$
or of $r_\mathrm{max}(t)$.

\section{Critical phenomena and non-universality}
Critical collapse for the Einstein-Vlasov
system has previously been studied numerically in \cite{RRS2,OC,RS}. 
By specifying an initial datum and then varying
its amplitude $A,$ a critical value $A_c$ of $A$ is found in the
sense that if $A>A_c$ the evolution of the initial datum will form
a trapped surface and collapse to a black hole, and if $A<A_c$ no
black hole will ever form. In the previous studies referred to
above it was stated that the solutions in the subcritical case
(i.e.\ $A<A_c$) will disperse to infinity. However, as we saw in
the previous section this is not quite true since there are (at
least) two possible scenarios: Either the solution will disperse
to infinity or it will be bound depending on the sign of the
binding energy $e_b$. Nevertheless, the value $A_c$ separates two
distinct behaviors, as was first discovered by Choptuik
\cite{Cho93} who studied the Einstein-scalar field system. From
his work a new topic developed which goes under the name of
critical phenomena, cf. \cite{G} for a review. Two types of matter
are distinguished in this context. Vlasov matter is said to be of
type~I whereas a scalar field and a perfect fluid are of type~II.
One fundamental difference between matter of type~I and type~II is
that the graph which describes the mass of the black hole (which
of course is zero if $A<A_c$) as a function of $A$ is
discontinuous at $A=A_c$ for type~I matter, and continuous for
type~II matter. For type~I matter there is thus a mass gap and
no possibility to get a black hole with arbitrarily small
mass as in the case of type~II matter. The characteristic
behavior of the critical solution itself has been carefully investigated
for type~II matter. A lot of evidence has been found
that it is self-similar and universal, where the latter property
means that it is independent of the initial datum, cf.\ \cite{G}. 
For type~I matter the
general conjecture is that the critical solution is static (and
unstable) and universal, see \cite{Cho93}. This conjecture has
been investigated in \cite{OC} and \cite{RS} for Vlasov matter,
and evidence that the critical solution is static was found. In
these investigations it is also claimed that some support for
universality is obtained, but it is concluded that further studies
need to be done. From the previous section it is clear that only
the static property, and not universality, can be a genuine
feature of the critical solution for Vlasov matter. Indeed, by
picking $f_0=f_s,$ where $f_s$ is the distribution function
of an unstable static solution we find $A_c=1$ as the critical 
amplitude and $f_s$ as the critical solution.
Since there are infinitely many unstable static solutions, and any
one will do as critical solution, universality is contradicted.
More precisely, within any family of steady states (which we have
investigated), specified by some choice of $k$ and $l$ in
(\ref{pol}), there are infinitely many unstable ones each of which
acts as a critical solution. And the class of steady states of the
form (\ref{pol}) is only a small subset of all possible forms of
steady states obtainable by the ansatz (\ref{ansats}), cf.\
\cite{RR4}.

If we turn to the other
conjectured property, that the critical solution is static, our
results strongly support that this is the case. We start from
subcritical initial data and tune the amplitude to a very high
degree of precision to get $A_c.$ We then find that in the
evolution the metric coefficients become more or less frozen 
after some time and stay so during a considerably long time
period before the solution starts to drift off outwards. The length
of the latter time period
depends on how carefully we have calibrated $A_c$. By a
considerably long time period we mean considerable in comparison to a typical
dynamical time scale, e.~g.\ a typical time to form a trapped
surface or the typical period time of an oscillation for a bound
state. This is further discussed in the next section on the error
estimates.

\section{The numerical errors}
To get an indication of the validity of our simulations we compute
at every time step the ADM mass $M(t)$ and the rest mass $M_0(t),$
which are conserved along true solutions of the system.
We define the errors
$e[M]$ and $e[M_0]$ by
\[
e[M](t)=\frac{|M(t)-M|}{M},\ e[M_0](t)=\frac{|M_{0}(t)-M_{0}|}{M_{0}},
\]
where $M=M(0)$ and $M_0=M_0 (0)$ are given by the initial datum.
As we have explained in previous sections, the evolution of an
initial datum will either form a trapped surface and collapse,
will be bound and never collapse, or it will disperse to infinity. 
In the latter two situations the errors in our simulations
are remarkably small as can be seen in the table below. On the other hand
when a trapped surface has formed a black hole singularity will develop 
(cf. \cite{DR}). The coordinates that we
are using are believed to avoid the singularity so that solutions will
be regular for all times. However, the solutions become
more and more peaked at a certain more or less frozen radius.
Of course a very fine grid is needed to resolve such sharp peaks.
We are using a fixed grid which does not change with time, and since
the solutions get more and more peaked as time goes on our grid is
not sufficient after a certain time of the collapse. This results in
an essential increase of the errors as seen in Table~5 in the case of
collapsing initial data. 

\begin{table}[h]
\centering
\begin{tabular}{|c|c|c|c|c|c|c|c|} \hline
$Z_c$ & $E_b$ & $A$ & TTS & PT & TF & $e[M](TF)$ & $e[M_0](TF)$ \\
\hline
$0.30$  & $0.039$ & 1.01 &  &    & 50  & 0.00035  & 0.000036   \\
$0.47$  & $0.039$ & 0.99 &   & 60  & 150  & 0.0013  & 0.000023   \\
$1.14$  & $-0.080$ & 0.99 &   &   & 20  & 0.00015  & 0.00011   \\
$1.14$  & $-0.080$ & 1.01 & 3.6 &   & 10  & 0.080  & 0.040   \\ \hline
\end{tabular}
\caption{Numerical errors}
\end{table}
For each of the simulations in Table~5 about $14\,000$ particles have 
been used.
This gives convenient running times of a few minutes,
but one can of course use a much
larger number of particles and for certain runnings we have used up to $10^6$
particles which gives running times of several hours
on a reasonably modern PC. 
To get an idea of the length of the final running time, $TF,$ we have
in the relevant cases included the time when a trapped surface forms, $TTS,$
and the period time, $PT,$ of an oscillation. In the collapsing
situation (row $4$) where the errors are of a different magnitude we also
computed the error at $t=2*TTS=7.2$ and found $e[M](7.2)=0.024$ and
$e[M_0](7.2)=0.046,$ in order to get an indication of the growth
of the errors. In conclusion, Table~5 clearly demonstrates that we
can keep $M$ and $M_0$ conserved
to a very high precision in the evolution as long as we are not considering
collapsing solutions. In the latter case the errors are still very
reasonable for a considerable running time compared to the time when a
trapped surface forms.

\end{document}